\documentclass[12pt]{article} % single column, double spaced

\usepackage{ol}
\usepackage{hyperref}
\usepackage{amsmath}
\usepackage{bm}% bold math

\begin{document}
\def\be{\begin{equation}}
\def\ee{\end{equation}}

% \twocolumn[ %% activate for two-column option

\title{Nonlinear absorption of ultrashort laser pulses in thin metal films}

\author{Giovanni Manfredi and Paul-Antoine Hervieux}

\address{Institut de Physique et Chimie des Mat{\'e}riaux de
Strasbourg, GONLO, BP 43, F-67034 Strasbourg, France}

\begin{abstract}Self-consistent simulations of the ultrafast
electron dynamics in thin metal films are performed. A regime of
nonlinear oscillations is observed, which corresponds to ballistic
electrons bouncing back and forth against the film surfaces. When
an oscillatory laser field is applied to the film, the field
energy is partially absorbed by the electron gas. Maximum
absorption occurs when the period of the external field matches
the period of the nonlinear oscillations, which, for sodium films,
lies in the infrared range. Possible experimental implementations
are discussed.
\end{abstract}


% ] %% activate for two-column option

\noindent The recent progress in the study of metallic
nanostructures is mainly due to the development of ultrafast
spectroscopy techniques, which allow the experimentalist to probe
the electron dynamics on a femtosecond (and, more recently,
attosecond) time scale. Typical ``pump-probe" experiments involve
perturbing the system via a first stronger pulse, followed by a
second weaker pulse that acts as a diagnostic tool. By modulating
the relative amplitude of the signals, as well as the delay
between the pump and the probe, it is possible to assess with
great precision the dynamical relaxation of the electron gas
\cite{Brorson,Suarez,Sun,Bigot}.

In the present work, we focus on the ultrafast electron dynamics
in thin metallic films. Several experiments have shown
\cite{Brorson,Suarez} that electron transport in thin metal films
occurs on a femtosecond time scale and involves ballistic
electrons traveling at the Fermi velocity of the metal $v_F$.
These findings were corroborated by accurate numerical simulations
\cite{MH}, which highlighted a regime of slow nonlinear
oscillations corresponding to ballistic electrons bouncing back
and forth on the film surfaces. These oscillations were recently
measured in transient reflection experiments on thin gold films
\cite{Liu}. The existence of this regime prompted us to analyze
the possibility of boosting energy absorption in the film by
optically exciting the electron gas at the frequency of the
nonlinear oscillations.

In the rest of this Letter, time is normalized in units of the
inverse plasmon frequency $\omega_{pe}^{-1}$, velocity in units of
the Fermi speed $v_{{F}}$, and length in units of $L_{F} =
v_{F}/\omega_{{pe}}$. For alkali metals we have $L_{{F}}=0.59
\left(r_s/a_0\right)^{1/2}$ \AA, $\omega_{pe}^{-1}=1.33\times
10^{-2} \left(r_s/a_0\right)^{3/2}$ fs, $E_{{F}}=
50.11\left(r_s/a_0\right)^{-2}$ eV and $T_F = 5.82 \times
10^{5}\left(r_s/a_0\right)^{-2}$ K, where $r_s$ is the
Wigner-Seitz radius. We concentrate primarily on sodium films, for
which $r_s=4a_0$ ($a_0=0.529~ {\rm \AA}$ is the Bohr radius).

We consider a system of electrons interacting via a Coulomb
potential and confined within a slab of thickness $L$. The ion
background is represented by a fixed density with soft edges,
$n_i(x)= \overline{n}_i\left[ 1+\exp\left( (|x|-L/2)/\sigma_i
\right) \right]^{-1}$, where $\overline{n}_i= 3/(4\pi r_s^3)$ is
the ion density of the bulk metal and $\sigma_i \ll L$ a
diffuseness parameter \cite{Calvayrac}. In this jellium model, the
self-consistent electrostatic potential depends only on the
coordinate normal to the surface (here noted $x$). Thus, the
motion of an electron parallel to the surface of the film is
completely decoupled from the motion normal to the surface and a
one-dimensional (1D) model can be adopted.

The electrons are initially prepared in a Fermi-Dirac equilibrium
at finite (but small) temperature. They are subsequently excited
by imposing a constant velocity shift $\Delta v_x=0.08v_F$ to the
initial distribution \cite{Calvayrac}. This scenario is
appropriate when no linear momentum is transferred parallel to the
plane of the surface (i.e., $q_{\parallel}=0$) and is relevant to
the excitation of the film with optical pulses \cite{Anderegg}.
For $q_{\parallel}=0$, only longitudinal modes (volume plasmon
with $\omega = \omega_{pe}$) can be excited.

After the excitation is applied, the electron distribution
function $f_e(x,v_x,t)$ starts evolving in time according to the
semiclassical Vlasov equation
\begin{equation}
\frac{\partial{f_{e}}}{\partial{t}} + v_x
\frac{\partial{f_{e}}}{\partial{x}} +\frac{e} {m_{e}}
\frac{\partial\phi}{\partial x} ~
\frac{\partial{f_{e}}}{\partial{v_x}} = 0, \label{vlasov}
\end{equation}
where $m_e$ is the electron mass and $e$ denotes the absolute
electron charge. The electrostatic potential is obtained
self-consistently, at each instant, from Poisson's equation
\begin{equation}
\frac{d^{2}\phi}{dx^2} =
\frac{e}{\varepsilon_0}[n_e(x,t)-n_i(x)]~, \label{poisson}
\end{equation}
with $n_e=\int f_{e} dv_x$. As a reference case, we studied a
sodium film with initial temperature $T_e=0.008 T_{{ F}}\simeq
300$ K, diffuseness parameter $\sigma_i=0.3 L_{\rm F}$, and
thickness $L=50L_{F} \simeq 59~$\AA \cite{Anderegg}.

\begin{figure}[htb]
\centerline{\includegraphics[width=8.cm,height=5.5cm]{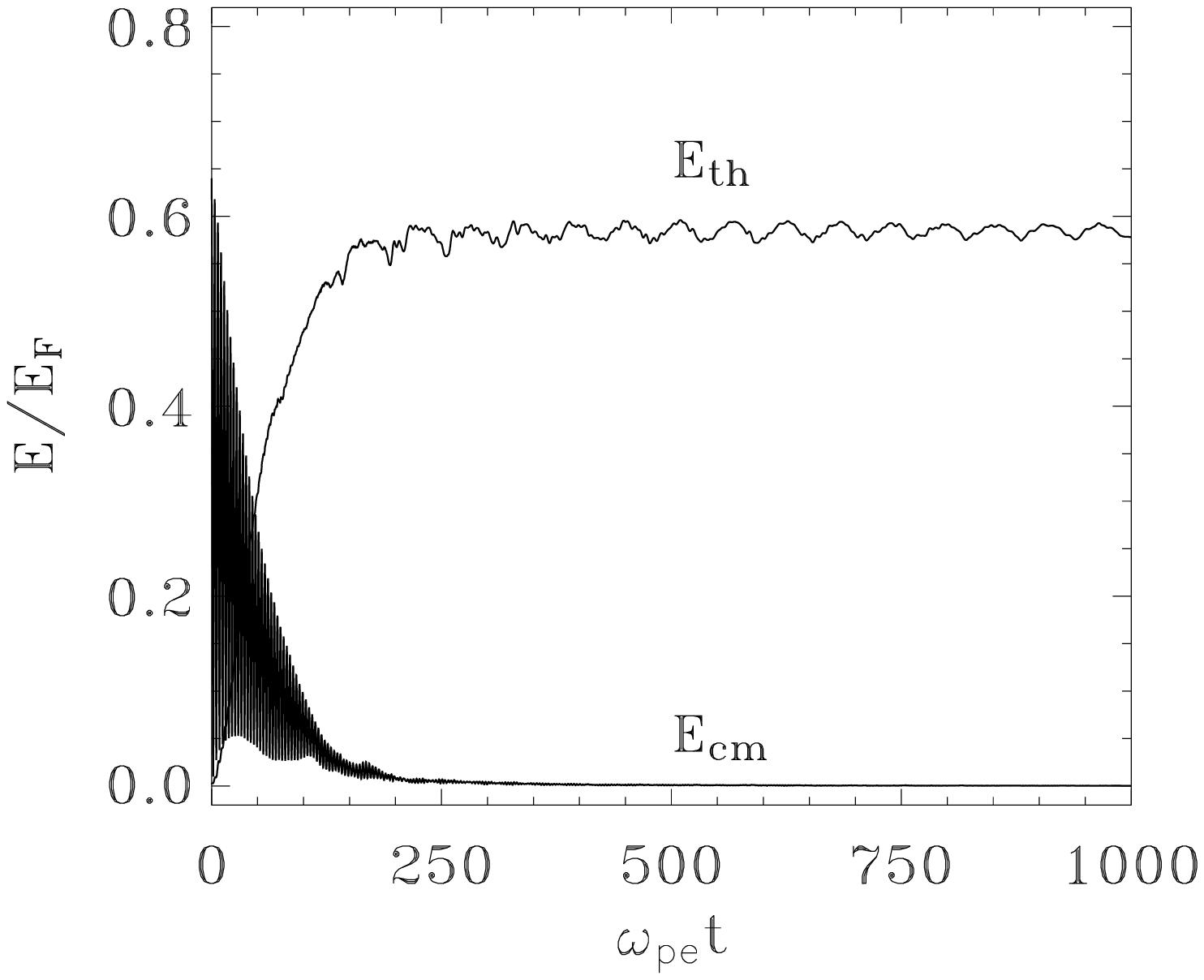}}
 \caption{\label{Fig:1}}
\end{figure}

The time evolution of the thermal $E_{\text{th}}$ and
center-of-mass $E_{\text{cm}}$ energies was analyzed \cite{MH}
(Fig. \ref{Fig:1}). During an initial rapidly-oscillating phase,
$E_{\text{cm}}$ is almost entirely converted into thermal energy
(Landau damping). After saturation, a slowly oscillating regime
appears, with period equal to $50 \omega_{pe}^{-1} \approx
5.3~\rm{fs}$. This period is close to the time of flight of
electrons traveling at the Fermi velocity and bouncing back and
forth on the film surfaces (further details are provided in our
previous work \cite{MH}).

The above nonlinear oscillations appear for all reasonable values
of the physical parameters. Preliminary studies suggest that
electron-electron collisions do not destroy this regime either, at
least for relatively low excitation energies and short times. It
is tempting, therefore, to investigate whether some kind of
resonant absorption can be achieved when the system is externally
excited at the same frequency of the nonlinear oscillations.

A similar scenario was investigated by Taguchi {\it at al.}
\cite{Taguchi} (building on an idea due to Brunel \cite{Brunel})
in order to simulate the interaction of an argon cluster with a
strong laser field ($\approx 10^{15}-10^{16}~{\rm W/cm^2}$). In
their simulations, the neutral cluster is quickly ionized by the
laser field, which heats the electrons up to 10~eV. At these
temperatures, the electrons behave classically and are initially
described by a Maxwell-Boltzmann distribution. In that case, the
electron transit velocity through the cluster is not clearly
defined and depends on the intensity of the laser (indeed, in the
classical case, there is no ``natural" oscillatory regime like the
one seen in Fig. 1). For a degenerate electron gas, the transit
velocity is unambiguously given by the Fermi velocity of the metal
and thus we expect an even neater resonance to occur.

Our conjecture can be tested in the following way. At time
$\omega_{pe} t = 1000$ ($\approx 106 ~{\rm fs}$ for a sodium
film), when the oscillatory regime is well established, we switch
on a small external electric field, uniform in space and
sinusoidal in time with period $T$: $E_{\rm ext} = E_0 \sin(2 \pi
t/T)$, where $E_0$ is the (constant) field amplitude. The
simulation is then continued with the external field on for
another $4000 \omega_{pe}^{-1} \approx 425 ~\rm{fs}$. This
situation corresponds to a laser pulse that is switched on very
quickly and lasts for a duration longer than $425 ~\rm{fs}$.

For an electron transit velocity exactly equal to $v_F$, we would
expect resonance for a laser period $T = 2L/v_F$ (= 100 in units
of $\omega_{pe}^{-1}$). The factor 2 comes from the fact that the
electric field must keep the same sign during a transit from one
surface to the other, and reverse sign during the ``return"
transit. We note that the resonance is expected to fall in the
infrared (IR) domain. Indeed, for a laser period $T = 100
\omega_{pe}^{-1} = 10.6~ \rm{fs}$, the corresponding wavelength is
$\lambda = 3.2 ~\mu\rm{m}$.

The amplitude $E_0$ of the laser field can be estimated by noting
that the total energy of the laser pulse is
$U = \left( \frac{1}{2} \varepsilon_0 E_0^2 \right) c \tau S$,
where $c$ is the speed of light in vacuum, $\tau$ is the pulse
duration, and $S$ is the surface of the laser spot. Typical values
for IR lasers \cite{Kaindl} are $S = 0.01 \textrm{mm}^2$ and $U=1
\mu \textrm{J}$, and by taking a pulse duration $\tau = 400
\textrm{fs}$ (similar to the duration used in the simulations), we
obtain an electric field $E_0 = 4.3 \times 10^{8}\textrm{V/m}$. In
the numerical simulations, the electric field is normalized to
$\overline{E} \equiv m_e v_F \omega_{pe}/e = 1.70 \times 10^{12}
(r_s/a_0)^{-5/2}~ \rm{V/m}$, yielding $\overline{E}= 5.31 \times
10^{10} \textrm{V/m}$ for sodium films. Therefore, by taking a
field amplitude $E_0= 0.01 \overline{E}$, we get a dimensional
value that is realistic for an IR laser pulse. This external field
amplitude is an order of magnitude smaller than the
self-consistent electric field present at the film surfaces.

\begin{figure}[htb]
\centerline{\includegraphics[width=8.cm,height=6cm]
{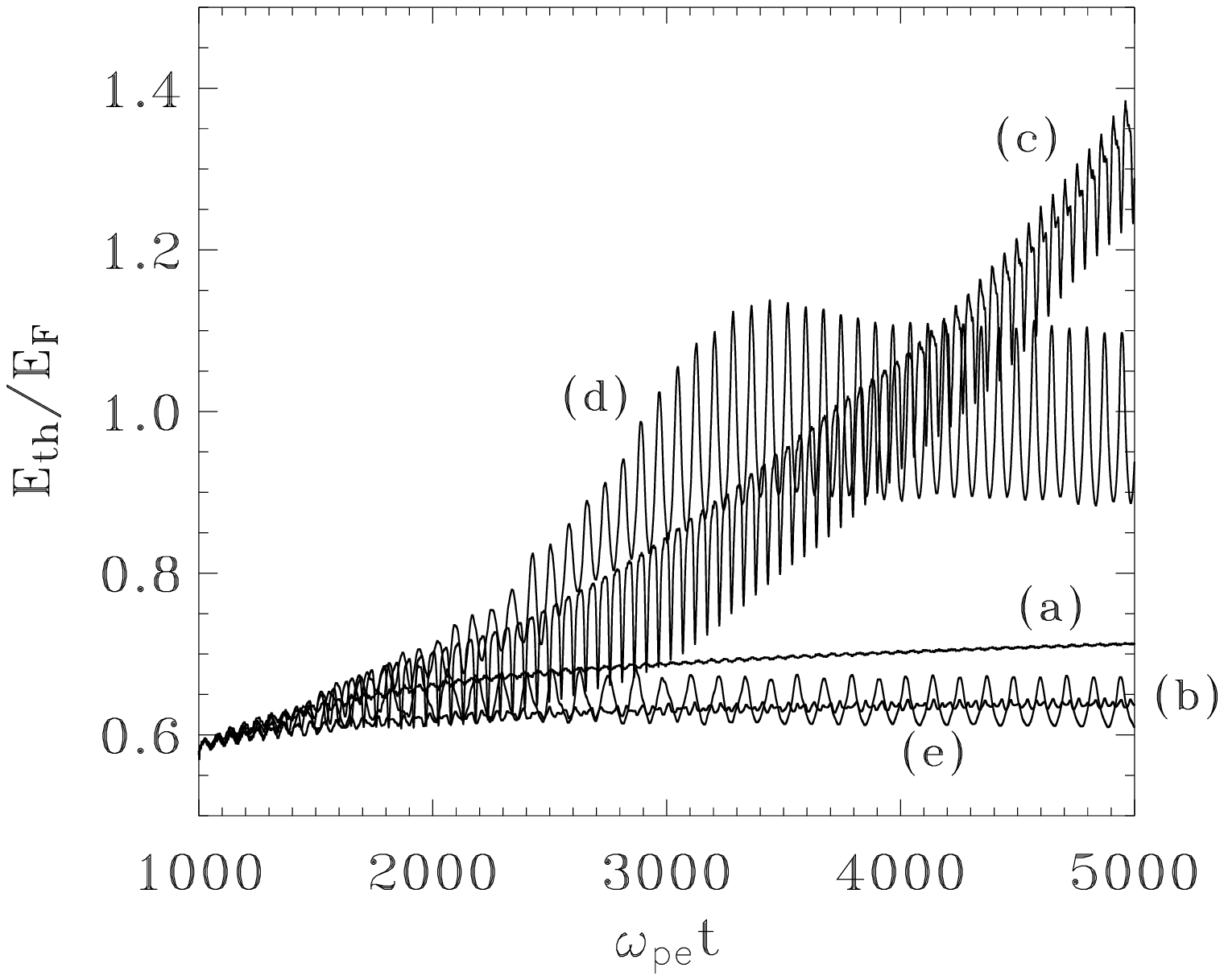}} \caption{\label{Fig:2}}
\end{figure}

The results for the reference case ($L=50 L_{F}$) are shown in
Fig. \ref{Fig:2}, where the electron thermal energy is plotted
against time. We observe that the absorption is clearly enhanced
for $\omega_{pe} T = 106$ and $\omega_{pe} T = 150$, whereas for
larger or smaller values virtually no energy is absorbed. We also
verified that the resonance does not depend on the phase of the
external oscillating field.

\begin{figure}[htb]
\centerline{\includegraphics[width=8.cm, height=5.5cm]
{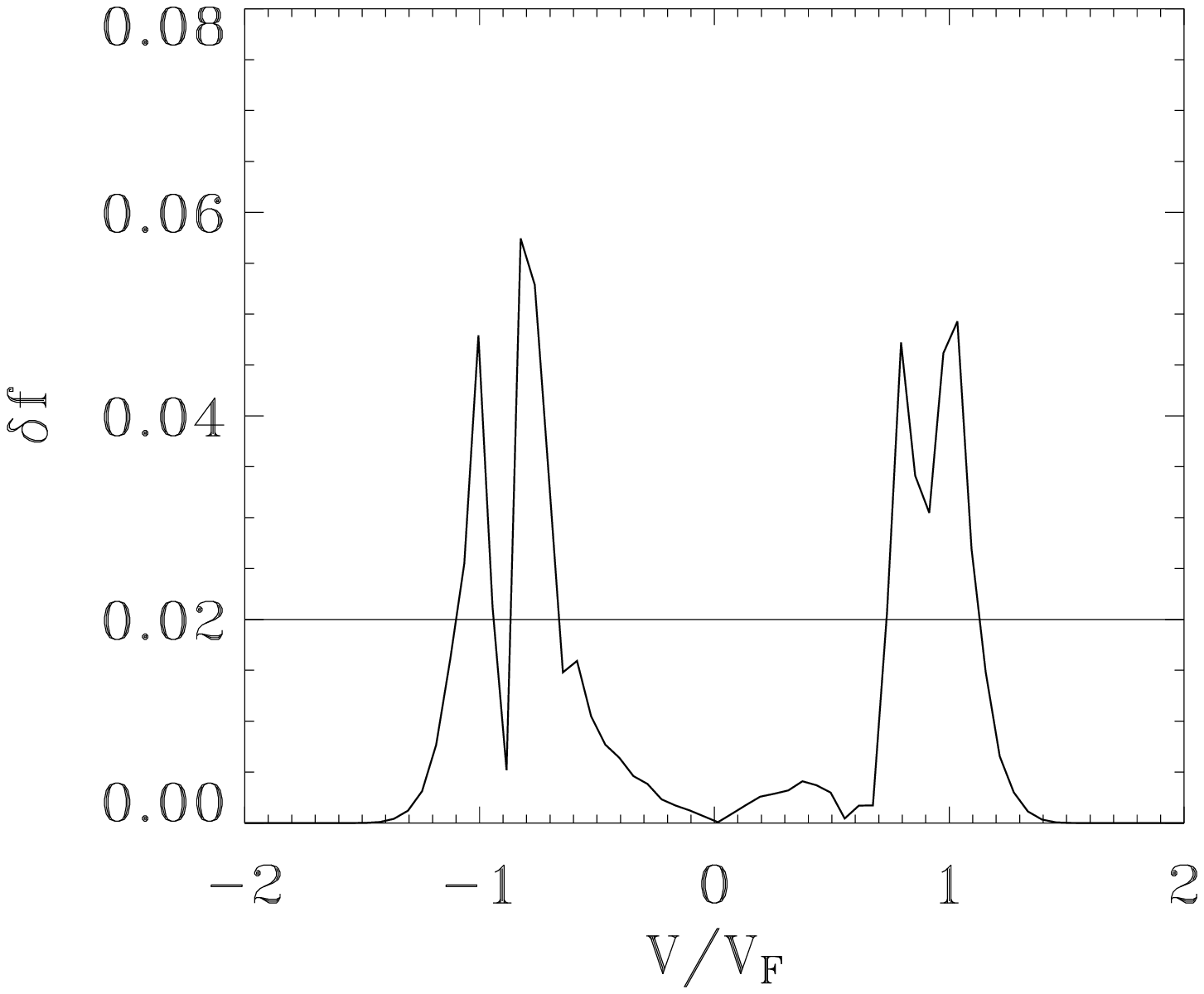}} \caption{\label{Fig:3}}
\end{figure}

The resonant period is close, but not exactly equal, to the
predicted value $\omega_{pe} T = 100$ and the resonance displays a
certain broadness. The latter can be explained by noting that a
certain dispersion exists in the electron velocities around $v_F$,
which generates a dispersion in the resonant period. If the period
is $T=2 L/v$, then the resonance broadness should be $|\delta T|
=(2 L/v^2)~\delta v$. In order to estimate the broadness, we plot,
in Fig. \ref{Fig:3}, the variation of the velocity distribution at
the center of the film:
$\delta f(v_x) = |f_e(x=0,v_x, \omega_{pe}t =1000) -
f_e(x=0,v_x,t=0)|$.
The distribution is indeed modified around the Fermi velocity, as
expected (see also Fig. 3 in Ref. 5). Note that $\delta f$ is not
symmetric around $v_x=0$, because the initial excitation was not
symmetric either. On closer inspection, the peaks occur at a
velocity slightly smaller (in absolute value) than $v_F$, roughly
$|v_x| \simeq 0.9 v_F$. Their broadness can be estimated by
assuming that a deviation of $2\%$ (relative to the maximum
$f_e=1$) is significant. Then, $v$ varies in the interval $0.7 v_F
<v < 1.1 v_F$ (and equivalently for negative velocities), so that
$\delta v \simeq 0.4 v_F$. This yields a broadness $\omega_{pe}
\delta T \simeq 50$, with $90 <\omega_{pe} T < 140$. This estimate
is compatible with the simulation results of Fig. 2, where the
resonance has clearly disappeared at $\omega_{pe} T = 73$ and 230.

\begin{figure}[htb]
\centerline{\includegraphics[width=8.cm, height=6cm]
{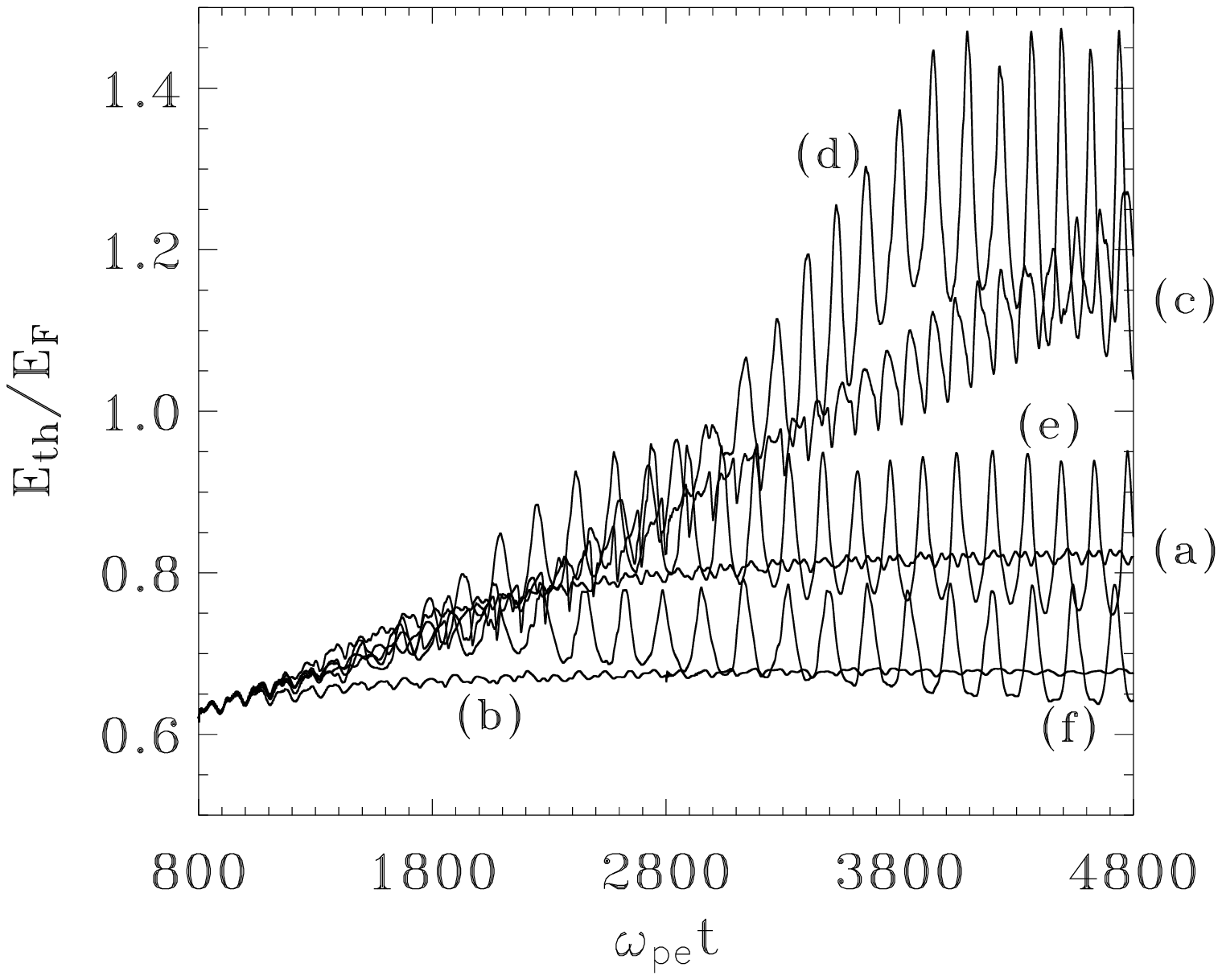}} \caption{\label{Fig:4}}
\end{figure}

In order to test the robustness of this nonlinear resonance
effect, we repeated the same numerical experiment with a thicker
film, $L=100 L_F$ (the initial evolution for this case is shown in
our previous work \cite{MH}). The resonant period is expected to
scale linearly with the film thickness, and indeed we observed
enhanced absorption for $\omega_{pe} T =212$ and 250 (Fig.
\ref{Fig:4}).

In contrast, we observed that the resonance virtually disappears
for thicker films, $L=200 L_F$ or larger. We interpret this result
by noticing that the existence of the resonance depends on
nonequilibrium electrons traveling coherently through the film.
The phase space portraits of the electron distribution function
(see Fig. 3 in our previous work \cite{MH}) show a complex
structure of traveling vortices. It is probable that, for
thicknesses larger than a certain threshold, the necessary
coherence is lost, so that the resonance cannot manifest itself.

The resonance also disappears for very small amplitudes of the
external field. For the reference case $L=50 L_F$, the resonance
is still observed for $E_0/\overline{E} =0.005$, but no longer for
$E_0/\overline{E} =0.001$. This may be related to the fact that
the absorbed energy at resonance $U_{\rm abs}$ scales
quadratically with the field amplitude: $U_{\rm abs} \propto e^2
E_0^2 T^2/m_e$ (this formula becomes exact for the harmonic
oscillator). For small fields, the resonance is thus very weak and
other factors (e.g., Landau damping) can easily erase it
completely.

In summary, we have shown the existence of a nonlinear absorption
regime in the electron dynamics of thin metal films. This effect
is generic and should not, in principle, depend on the nature of
the metal. The resonance occurs in the IR domain and should be
accessible via experiments employing ultrafast laser sources with
standard specifications. This absorption mechanism could be used
as an optical diagnostic technique to determine, for instance, the
thickness of the film, or to obtain information on the electronic
distribution.

We would like to thank J.-Y. Bigot for initiating this project and
providing constant support. We also thank V. Halt{\'e} and J. L{\'e}onard
for their helpful comments. The numerical calculations were
performed on the computers of the IDRIS computing center in Orsay,
France.
\bigskip

%\pagebreak

\newpage

\newpage

\newpage

\newpage

\newpage

{\bf \large FIGURE CAPTIONS}
\bigskip

{\bf Fig. 1:}  Initial time evolution of the thermal and
center-of-mass energies for a film thickness $L=50L_F$.

\bigskip
{\bf Fig. 2:} Time evolution of the thermal energy in the presence
of an external electric field. The external field is switched on
at $\omega_{pe} t=1000$. (a) $\omega_{pe} T = 27$; (b)
$\omega_{pe} T = 73$; (c) $\omega_{pe} T = 106$; (d) $\omega_{pe}
T = 150$; (e) $\omega_{pe} T = 230$. The results are for a film of
thickness $L=50L_F$.

\bigskip
{\bf Fig. 3:} Variation of the electron velocity distribution with
respect to the initial Fermi-Dirac equilibrium, at the center of
the film, at time $\omega_{pe} t =1000$.

\bigskip
{\bf Fig. 4:} Same as Fig. \ref{Fig:2} for a film of thickness
$L=100L_F$. (a) $\omega_{pe} T = 90$; (b) $\omega_{pe} T = 150$;
(c) $\omega_{pe} T = 212$; (d) $\omega_{pe} T = 250$; (e)
$\omega_{pe} T = 290$; (f) $\omega_{pe} T = 350$.


\begin{thebibliography}{99}
\bibitem{Brorson} S. D. Brorson, J. G. Fujimoto, and E. P. Ippen, Phys. Rev. Lett. {\bf 59},
1962 (1987).
\bibitem{Suarez} C. Su{\'a}rez, W. E. Bron, and T. Juhasz, Phys. Rev. Lett. {\bf 75},
4536 (1995).
\bibitem{Sun} C.-K. Sun, F. Vall\'{e}e, L. H. Acioli, E. P. Ippen and
J. G. Fujimoto, Phys. Rev. B {\bf 50}, 15337 (1994).
\bibitem{Bigot} J.-Y. Bigot, V. Halt\'{e}, J.-C. Merle, and A. Daunois, Chem. Phys. {\bf 251}, 181 (2000).
\bibitem{MH} G. Manfredi and P.-A. Hervieux, Phys. Rev. B {\bf 70},
201402(R) (2004).
\bibitem{Liu}
X. Liu, R. Stock, and W. Rudolph, CLEO/IQEC and PhAST Technical
Digest on CDROM (The Optical Society of America, Washington, DC,
2004), IWA4.
\bibitem{Calvayrac} F. Calvayrac, P.-G. Reinhard, E. Suraud,
and C. Ullrich, Phys. Rep. {\bf 337}, 493 (2000).
\bibitem{Anderegg} M. Anderegg, B. Feuerbacher, and B. Fitton,
Phys. Rev. Lett. {\bf 27}, 1565 (1971).
\bibitem{Taguchi} T. Taguchi, T. M. Antonsen, Jr., and H. M.
Milchberg, Phys. Rev. Lett. {\bf 92}, 205003 (2004).
\bibitem{Brunel}F. Brunel, Phys. Rev. Lett. {\bf 59}, 52 (1987).
\bibitem{Kaindl} R. A. Kaindl, M. Wurm, K. Reimann, P. Hamm, A. M.
Weiner, M. Woerner, J. Opt. Soc. Am. B {\bf 17}, 2086 (2000).

\end{thebibliography}
\end{document}